\documentclass[12pt,preprint]{aastex}

\newcommand{\kms}{\ensuremath{\rm km\,s^{-1}}}
\newcommand{\ms}{\ensuremath{\rm m\,s^{-1}}}
\newcommand{\gcmc}{\ensuremath{\rm g\,cm^{-3}}}

\newcommand{\teff}{\ensuremath{T_{\rm eff}}}
\newcommand{\logg}{\ensuremath{\log{g}}}
\newcommand{\vsini}{\ensuremath{v \sin{i}}}
\newcommand{\feh}{[Fe/H]}

\newcommand{\rsun}{\ensuremath{R_\sun}}
\newcommand{\msun}{\ensuremath{M_\sun}}
\newcommand{\lsun}{\ensuremath{L_\sun}}

\newcommand{\rstar}{\ensuremath{R_\star}}
\newcommand{\mstar}{\ensuremath{M_\star}}
\newcommand{\loggstar}{\ensuremath{\logg_\star}}
\newcommand{\lstar}{\ensuremath{L_\star}}

\newcommand{\rhostar}{\ensuremath{\rho_\star}}

\newcommand{\rpl}{\ensuremath{R_{\rm P}}}
\newcommand{\mpl}{\ensuremath{M_{\rm P}}}

\newcommand{\rhopl}{\ensuremath{\rho_{\rm P}}}
\newcommand{\loggpl}{\ensuremath{\logg_{\rm P}}}
\newcommand{\teq}{\ensuremath{T_{\rm eq}}}

\newcommand{\rjup}{\ensuremath{R_{\rm J}}}
\newcommand{\mjup}{\ensuremath{M_{\rm J}}}

\newcommand{\koicur}{Kepler-5}
\newcommand{\koicurb}{Kepler-5b}

\newcommand{\koicurCCra}{\ensuremath{19^{\mathrm{h}}57^{\mathrm{m}}37^{\mathrm{s}}.68}}
\newcommand{\koicurCCdec}{\ensuremath{+44^{\circ}02'06''.2}}
\newcommand{\koicurCCkic}{KIC~8191672}
\newcommand{\koicurCCtwomass}{2MASS~19573768+4402061}

\newcommand{\koicurLCar}{\ensuremath{6.06\pm0.14}}		
\newcommand{\koicurLCrprstar}{\ensuremath{0.08195^{+0.00030}_{-0.00047}}}	
\newcommand{\koicurLCimp}{\ensuremath{0.393^{+0.051}_{-0.043}}}		
\newcommand{\koicurLCi}{\ensuremath{86\fdg3\pm0.5}}		%

\newcommand{\koicurLCP}{\ensuremath{3.548460\pm0.000032}}	

\newcommand{\koicurLCT}{\ensuremath{2454955.90122\pm0.00021}}	

\newcommand{\koicurSMEteff}{\ensuremath{6297\pm60}}	
\newcommand{\koicurSMEfeh}{\ensuremath{+0.04\pm0.06}}	
\newcommand{\koicurSMElogg}{\ensuremath{3.96\pm0.10}}	
\newcommand{\koicurSMEvsin}{\ensuremath{4.8\pm1.0}}	

\newcommand{\koicurYYmlong}{\ensuremath{1.374^{+0.040}_{-0.059}}}	%
\newcommand{\koicurYYrlong}{\ensuremath{1.793^{+0.043}_{-0.062}}}	%
\newcommand{\koicurYYlogg}{\ensuremath{4.07\pm0.02}}			%
\newcommand{\koicurYYlum}{\ensuremath{4.67^{+0.63}_{-0.59}}}		%
\newcommand{\koicurYYage}{\ensuremath{3.0\pm0.6}}			%

\newcommand{\koicurRVK}{\ensuremath{227.5\pm2.8}}			
\newcommand{\koicurRVgamma}{\ensuremath{0}}				
\newcommand{\koicurRVmean}{\ensuremath{-46.7\pm4.1}}			

\newcommand{\koicurPPlogg}{\ensuremath{3.41\pm0.03}}			%
\newcommand{\koicurPParel}{\ensuremath{0.05064\pm0.00070}}			
\newcommand{\koicurPPrho}{\ensuremath{0.894\pm0.079}}			%

\newcommand{\koicurPPm}{\ensuremath{2.114^{+0.056}_{-0.059}}}		%
\newcommand{\koicurPPr}{\ensuremath{1.431^{+0.041}_{-0.052}}}		%
\newcommand{\koicurPPteq}{\ensuremath{1868\pm284}}			%

\shorttitle{\koicur}
\shortauthors{Rowe et al.}
\slugcomment{Version 2.5 --- \today}

\begin{document}

\title{Discovery of the Transiting Planet \koicurb}

\author{
David~G.~Koch,\altaffilmark{1}
William~J.~Borucki,\altaffilmark{1}
Jason~F.~Rowe,\altaffilmark{1,2}
Natalie~M.~Batalha,\altaffilmark{3}
Timothy~M.~Brown,\altaffilmark{4}
Douglas~A.~Caldwell,\altaffilmark{5}
John.~Caldwell,\altaffilmark{6}
William~D.~Cochran,\altaffilmark{7}
Edna~DeVore,\altaffilmark{5}
Edward~W.~Dunham,\altaffilmark{8}
Andrea~K.~Dupree,\altaffilmark{9}
Thomas~N.~Gautier~III,\altaffilmark{10}
John~C.~Geary,\altaffilmark{9}
Ron~L.~Gilliland,\altaffilmark{11}
Steve~B.~Howell,\altaffilmark{12}
Jon~M.~Jenkins,\altaffilmark{5}
David~W.~Latham,\altaffilmark{9}
Jack~J.~Lissauer,\altaffilmark{1}
Geoff~W.~Marcy,\altaffilmark{13}
David~Morrison,\altaffilmark{1}
Jill~Tarter,\altaffilmark{5}
}

\altaffiltext{1}{NASA Ames Research Center, Moffett Field, CA 94035}
\altaffiltext{2}{NASA Postdoctoral Program Fellow}
\altaffiltext{3}{San Jose State University, San Jose, CA 95192}
\altaffiltext{4}{Las Cumbres Observatory Global Telescope, Goleta, CA 93117}
\altaffiltext{5}{SETI Institute, Mountain View, CA 94043}
\altaffiltext{6}{York University, Toronto, Ontario, Canada}
\altaffiltext{7}{University of Texas, Austin, TX 78712}
\altaffiltext{8}{Lowell Observatory, Flagstaff, AZ 86001}
\altaffiltext{9}{Harvard-Smithsonian Center for Astrophysics, Cambridge, MA 02138}
\altaffiltext{10}{Jet Propulsion Laboratory/California Institute of Technology, Pasadena, CA 91109}
\altaffiltext{11}{Space Telescope Science Institute, Baltimore, MD 21218}
\altaffiltext{12}{National Optical Astronomy Observatory, Tucson, AZ 85719}
\altaffiltext{13}{University of California, Berkeley, Berkeley, CA 94720}



\begin{abstract}
We present 44 days of high duty cycle, ultra precise photometry of the
13th magnitude star \koicur\ (KIC~8191672, \teff=6300 K, \logg=4.1), which exhibits periodic
transits with a depth of 0.7\%.  Detailed modeling of the transit is
consistent with a planetary companion with an orbital period of
$\koicurLCP$ days and a radius of $\koicurPPr$ \rjup.  Follow-up
radial velocity measurements with the Keck HIRES spectrograph on 9
separate nights demonstrate that the planet is more than twice as massive as Jupiter with a mass of $\koicurPPm$ \mjup\ and a mean density of $\koicurPPrho$ g/cm$^3$.

\end{abstract}

\keywords{planetary systems --- stars: individual (\koicur,
\koicurCCkic, \koicurCCtwomass) --- techniques: spectroscopic --- Facilities: \facility{The Kepler Mission}.}

\section{INTRODUCTION}\label{intro}

The launch of {\it Kepler} offers a special opportunity to
study the nature of transiting extrasolar planets through transit
photometry.  The {\it Kepler} mission is designed to determine the frequency
of terrestrial-size planets. The {\it Kepler} mission also
provides an excellent platform to study and understand the wide
variety of extrasolar planets.  From the spacecraft's Earth-trailing orbit
heliocentric photometric measurements are devoid of artifacts usually associated
with instruments in close proximity to the Earth, due to problems such
as the day-night cycle and effects imposed by the Earth's atmosphere
on ground based observations, and orbital effects for satellites in low
earth orbit.

We report the discovery and confirmation of \koicurb, a strongly
irradiated transiting hot Jupiter.  We describe the {\it Kepler} photometry
and transit modeling, and the follow-up observations used to confirm
that \koicurb\ is a planet, including an orbital solution using radial
velocities obtained with HIRES on Keck 1.


\section{{\it KEPLER} PHOTOMETRY}

The {\it Kepler} photometer consists of a 0.95-m Schmidt telescope feeding an array of 42 CCDs with a broad (430 - 890 nm) spectral response \citep{koc10}.  From its Earth-trailing heliocentric orbit and single field of view, {\it Kepler} can monitor more than 100,000 stars nearly continuously for the lifetime of the mission.

Photometry of \koicurb\ (\koicurCCkic, $\alpha = \koicurCCra$, $\delta = \koicurCCdec$, Kepmag=13.369) was obtained during 1 May - 15 June 2009, at an integration time of 29.426 minutes for each observation.  This run includes 9.7 days of photometric data taken while ground analysis was being performed in preparation for start of science operations, and is referred to as the Q0 data set \citep{has10}. Kelper science operations began on 12 May 2009, which compose the second data set of 33.5 days of photometry referred to as the Q1 data set. There is a gap with a length of 30 hours between Q0 and Q1 to facilitate contact between the spacecraft and groundstations for downlink of the Q0 data stream.

\begin{figure}
\begin{center}
\includegraphics[scale=0.9]{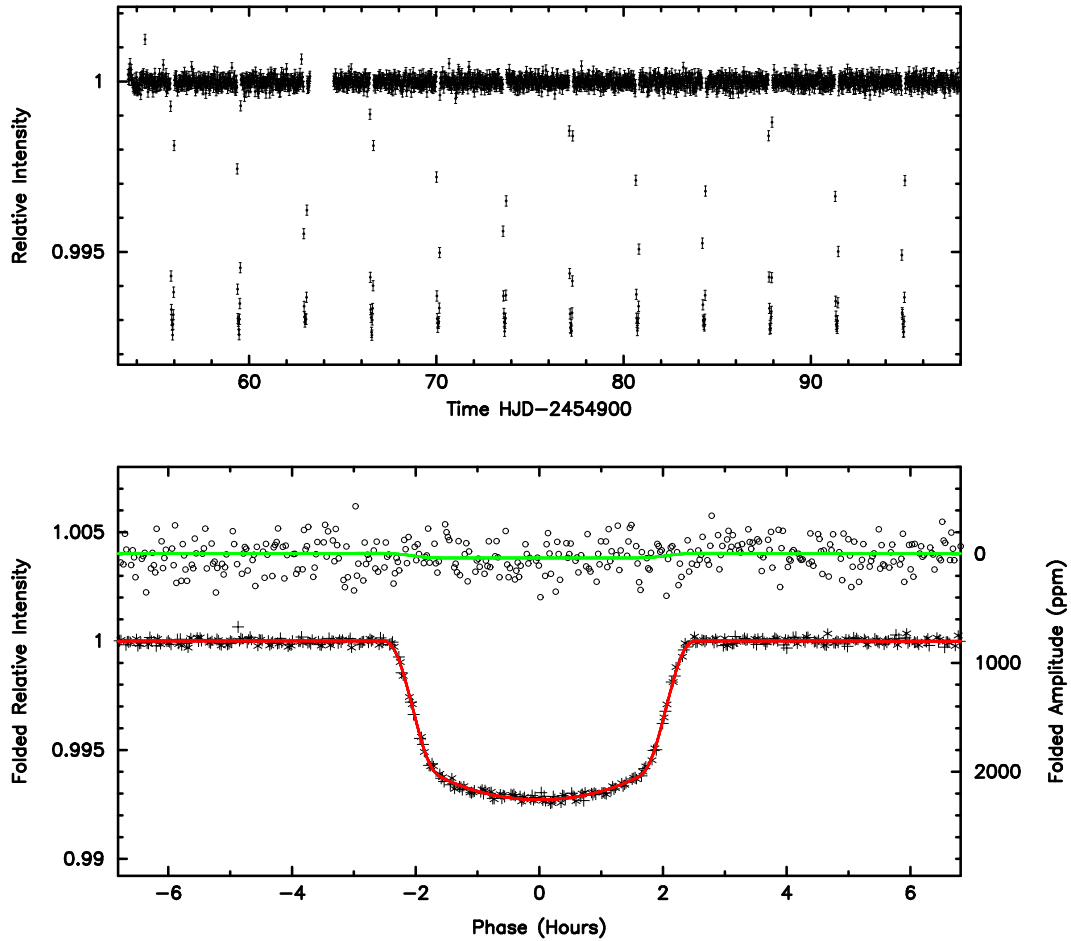}
\end{center}
\caption[\koicur\ Photometry]{The phased light curve of \koicur\ 
containing 12 transits observed by {\it Kepler} between 1 May and 15
June 2009.  The upper panel shows the full 44-day time series after
detrending.  The bottom panel shows the light curve folded with the
orbital period.  The lower curve shows the primary eclipse, with the
fitted transit model overplotted in red and corresponding scale to the
left. The upper curve covers the expected time of occultation,
with the fitted model overplotted in green and corresponding scale
found to the right, expanded by a factor of 5 relative to the scale
for the transit data.}
\label{fig:koi18phot}
\end{figure}

After pipeline data processing and photometry extraction, as described
in \citet{jen10} and \citet{cal10}, the time series stream was
detrended with a running 1-day mean.  All observations that occurred
during a transit were rejected from the evaluation of the mean.  The
detrended photometry is presented in Figure~\ref{fig:koi18phot}.

\subsection{Initial Transit Fits and False-Positive Detection}\label{tfit}

From examination of Q0 light curves, \koicur\ and a few dozen other candidates were initially flagged as {\it Kepler} Objects of Interest (KOI).  To identify the best candidates for ground-based follow-up observations, a number of metrics are computed to help recognize and reject stellar binary systems that can mimic a planetary transit signal. A quick summary of these steps, as it relates to \koicur\ are listed below.  For a full description of the steps taken to identify false positives see \citet{bat10}.

The transit lightcurve is modeled using the analytic expressions of \citet{man02} using $V$-band non-linear limb darkening parameters \citet{cla00}.  The stellar radius (\rstar), effective temperature ($\teff$) and surface gravity ($\logg$) are fixed to values adopted from {\it Kepler} Input Catalog (KIC).  The stellar mass ($\mstar$) is usually calculated from \logg\ and \rstar.  In the case of \koicur, a larger stellar radius of 1.9 \rsun\ was required to be consistent with the transit duration.  With \mstar\ and \rstar\ fixed to their initial values, a transit fit is then computed to determine the orbital inclination, planetary radius, and depth of the occultation (passing behind the star) assuming a circular orbit.  The best fit is found using a Levenberg-Marquardt minimization algorithm \citep{pre92}.

These initial fits are used to determine whether the transit is representitive of a planetary event.  In the case of a stellar binary where the surface brightness of the two components differ, the observed depth of the odd and even numbered transits will differ.  There was no measurable difference in the transit depths to report.



For systems that show only primary transits, a search is made for a weak occultation assuming that the orbit is circular, namely at phase 0.5.  In the cases where an occultation is found with significance greater than $2\sigma$ for the depth, the dayside temperature of the planet can be estimated.  From the depth of the occultation, the flux ratio of the planet and star ($F_{\rm P}/F_{\star}$) over the instrumental bandpass can be obtained. The depth of the transit indicates the ratio of the planet and star radii.  By assuming that the star and planet both behave as blackbodies and the flux ratio is bolometric, the dayside effective temperature can be estimated.  For \koicur\ we find $F_{\rm P}/F_{\star} = 3.6 \times 10^{-5}$.  This gives a planetary effective temperature of $\teff = 1720\pm214$\,K, where an error of 30\% is assumed for the input stellar luminosity and radius.  This estimate is a lower limit, as a significant fraction of the planetary flux is emitted at wavelengths longer than the red edge of the {\it Kepler} bandpass, but it is a useful diagnostic in order to determine whether the depth of the occultation is consistent with a strongly irradiated planet.  To make this comparison we can estimate the equilibrium temperature,
\begin{equation}\label{eq:teq}
T_{\rm eq} = T_{\star}(R_{\star}/2a)^{1/2} [f(1-A_{\rm B})]^{1/4},
\end{equation}
for the companion, where $R_{\star}$ and $T_{\star}$ are the stellar radius and temperature, with the planet at distance $a$ with a Bond albedo of $A_{\rm B}$, and $f$ is a proxy for atmospheric thermal circulation.  We assume $A_{\rm B} = 0.1$ for highly irradiated planets \citep{row06} and $f=1$ for efficient heat distribution to the night side. Assuming stellar irradiation is the primary energy source, these parameter choices give a rough estimate for the dayside temperature of the planet. Assuming a 30\% error in the input stellar parameters and that the star and planet act as blackbodies we find $T_{eq}=1810\pm289$ for \koicurb.  The consistency of \teff\ and $T_{eq}$ to first order suggests that the occultation is consistent with a strongly irradiated planet.  If the estimate of \teff\ were found to be much larger than $T_{eq}$, then the companion is likely to be self-luminous and is probably a star.
 
\subsection{Centroid Shifts}\label{centroids}

The {\it Kepler} pipeline provides measurements of the centroids of stellar images with a
precision of about 0.1 millipixel for the brightness of \koicur\
(Batalha, 2010).  Examination of the image centroids during transits
revealed shifts with an amplitude of 0.2 millipixels.  Such a motion
can result if the photo-aperture for \koicur\ includes light from
another source.  We examine and demonstrate with speckle and
adaptive-optic (AO) images in Section~\ref{AO} that the centroid motion
observed for \koicur\ can be accounted for by the presence of nearby
faint stars with constant brightness.

\section{FOLLOW-UP OBSERVATIONS}

After a KOI passes the above tests, additional ground-based follow-up
observations are obtained as described in \citet{gau10}.  These
observations include high-resolution imaging to search for additional
sources of flux within the {\it Kepler} photometric aperture that would
dilute the depth of the transit, and reconnaissance spectroscopy to
confirm and refine the KIC stellar classification and to search for
evidence of stellar companions.

\subsection{High-Resolution Imaging}\label{AO}

Ground-based visible-light speckle imaging from the WIYN Telescope and
near-infrared AO imaging from the Mt. Palomar 5-m telescope show that
\koicur\ has two companions. The first companion is $0.9\arcsec$ away
and 5.2 magnitudes fainter, as seen in the NIR-AO image, but is not
visible in the speckle image.  The other companion, KIC~8191680, is
$7.3\arcsec$ away with a {\it Kepler} magnitude of 17.69. The difference in
{\it Kepler} magnitude between \koicur\ and KIC~8191680 can be used to
compute the expected centroid shift, depending on which object is
assumed to be the source of the transit signal.

If KIC~8191680 is a background eclipsing binary, then the centroids
are expected to shift by +14.0 and $-0.8$ millipixels in the column
and row directions of the detector, respectively.  If KIC~8191680
shows no flux changes and \koicur\ is indeed the source of the transit
signal, then the expected centroid shift is $-0.3$ and 0.0
millipixels, which is consistent with the centroid shift reported in
Section~\ref{centroids}.  The transit depth is estimated to be diluted by
$\sim 2\pm0.2$\%.  We include the effects of dilution in our transitfits.

The speckle image shows a single star within its $2\arcsec$ square
field. This rules out very close background eclipsing binaries that might
simulate the observed transits, except for stars lying closer than
about $\sim 0.1\arcsec$ from the target star.

\subsection{Reconnaissance Spectroscopy}

The FIbre-fed Echelle Spectrograph (FIES) on the 2.5-m Nordic Optical
Telescope (NOT) was used on 4 June 2009 to obtain a spectrum of
\koicur\ for the purposes of stellar classification.  Stellar
parameters of $\logg = 4.0$ and $\teff = 6500$\,K were estimated,
which were used to refine the transit model fits described in
\S\ref{tfit}.  The spectrum shows no indication that the system includes an
eclipsing binary or has a composite spectrum.

\subsection{HIRES Spectroscopy}

\begin{figure}
\begin{center}
\includegraphics[scale=0.5]{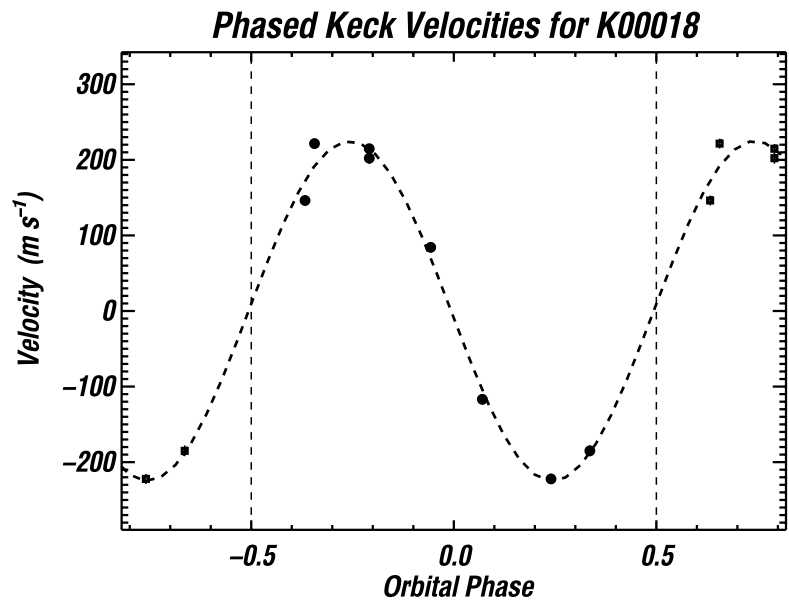}
\end{center}
\caption[\koicur\ Radial Velocities]{The phased radial-velocity curve 
for \koicur\ consisting of 8 epochs observed using the Keck/HIRES
spectrometer spanning 126 days.  The overplotted fit assumes a circular
orbit, phased to match the transit photometry.  A tabulation of the
radial-velocity measurements used for this Figure may be found in the
online materials.}
\label{fig:koi18rv}
\end{figure}

HIRES spectra were obtained from 3-6 June, 2-4 July, and on 6 October,
2009.  Figure~\ref{fig:koi18rv} shows the radial velocities for
\koicur\ folded with the photometric period of the planet. An
analysis of the Keck/HIRES template spectrum by D.~Fischer using SME
\citep{Valenti:96} measured the stellar parameters as listed in 
Table~\ref{tab:parameters}.  A fit to the velocities with the
eccentricity fixed to zero measures a reflex velocity semi-amplitude
of $K = 228\pm8\,\ms$ with residuals of $15\,\ms$. A bisector analysis
of the HIRES velocities does not show any significant variations
correlated with the radial velocities.

\section{ANALYSIS}

The {\it Kepler} photometric and HIRES radial-velocity measurements can be simultaneously modeled.  The data is fit for the center of transit time, period, impact parameter ($b$), the scaled planetary radius (R$_{\rm p}$/R$_{\star})$, the amplitude of the radial velocity (K), photometric and velocity zeropoints and $\zeta$/R$_{\star}$.  The last term, $\zeta$/R$_{\star}$ is related to the transit duration ($T_d=2(\zeta/\rstar)^{-1}$) and the mean stellar density \citep{pal09}.  We also account for non-circular orbits by modeling for e cos($\omega$), e sin($\omega$), where e is the orbital eccentricity and $\omega$ is argument of periastron.


The modeled stellar density is strongly dependent on the impact parameter,
$b$, of the planetary orbit.  It was discovered that the adopted
limb-darkening parameters were placing unrealistic constraints on $b$.
Specifically, the best $\chi^2$ would occur at $b=0$ for a
majority of the models of {\it Kepler} light curves.  The model fits were
repeated for various choices of limb darkening, including values
generated specifically for the {\it Kepler} bandpass with Atlas 9 models
by A.~Prsa.  Detailed examination of the fits revealed that the
limb darkening models were over-predicting the curvature of the base of
the transit shape.  To alleviate this problem, limb
darkening coefficients were computed for the three known exoplanet
systems in the {\it Kepler} field of view, TrES-2 \citep{dae09}, HAT-P-7 
\citep{pal08,gil10}, and HAT-P-11 \citep{bak09}, based on fits fixed to the
published values of \mstar, \rstar, and $i$.  We then linearly
interpolate between these values for candidates with different
temperatures.  It is estimated this procedure results in a $\sim
0.5^{\circ}$ systematic error on the quoted value of $i$ reported in
Table~\ref{tab:parameters}.

The error distribution for the stellar densities that fit the transit light
curve are obtained from a Markov-Chain-Monte-Carlo (MCMC) analysis.  We adopt the approach outlined in \S4.3 of \citet{for05}.  Convergence was tested by generating 10 Markov-chains each with $10^6$ steps and different initial conditions.  We checked that each chain had statistically similar means and distributions.  Given the stellar effective temperature from the Keck/HIRES template and the mean stellar density from transit photometry we employ the \rhostar\ method.  This method uses stellar evolution Yale-Yonsei tracks to calculate the range of stellar parameters consistent with the observations.  A full description of the \rhostar\ method is described in citet{bor10b}.   

The output of the \rhostar\ method is a set of Markov Chains giving a consistent set of $10^5$ allowed \mstar\ and \rstar\ pairs.  The distribution of stellar mass and radius allows us to define a transition probability to compute a MCMC analysis to determine the most likely model values.  We fit for stellar mass and radius, planetary mass and radius, center time of transit, orbital period and inclination, depth of the occulation and allow for non-circular orbits through the inclusion of e cos($\omega$) and e sin($\omega$).
We generated 10 sets of Markov-Chains with $10^6$ elements and use all 10 sets to generate distributions for the model fits.  

The errors in Table~\ref{tab:parameters} are centered around the mode of each model parameter which represents the value of highest relative probability.  The mode is calculated by binning the bootstrapped parameters into 50 equally-spaced bins.  We chose to use the mode to determine the most likely system parameters as some distributions, such as the stellar mass, are bi-modal. We report the bounds that enclose $\pm68\%$ of the
samples centred on the mode and report these values in Table~\ref{tab:parameters}.

\section{DISCUSSION}

Similar to {\it Kepler}-7 \citep{lat10}, the \koicur\ host star is not much hotter than the Sun, $\teff = \koicurSMEteff$, but is much more massive and larger than the Sun, $\mstar = \koicurYYmlong\,\msun$ and $\rstar = \koicurYYrlong\,\rsun$.  With a mean stellar density of  $\rhostar = 0.33\pm0.02$\ \gcmc\ the star is on its way to becoming a shell-burning subgiant.  Similar to {\it Kepler-4} \citep{bor10b} and {\it Kepler}-7 there are two distinct sets of model parameters that fit the observations equally well.  There are two peaks in the mass distribution centered at $\mstar = 1.21$ and $1.38\,\msun$.  While our analysis favours the larger mass, the lower mass choice can not be ruled out.

The planet is found to have a mass of $\mpl = \koicurPPm\,\mjup$ and a radius of $\rpl = \koicurPPr\,\rjup$, where \rjup\ refers to the equatorial radius of Jupiter.  This makes the planet more massive and larger than Jupiter.  The mean planetary density is $0.89\pm0.08$\ \gcmc, which is not uncommon amongst the known population of transiting Jupiter-sized planets.  Figure 3 from \citet{lat10} compares the mass and radius of \koicurb\ to other known transiting extrasolar planets.

The photometry points to a weak (2$\sigma$) detection of an occultation.  As correlations in the timeseries may lead to a false detection of an occultation we computed a Fourier Transform and autocorrelation of the dataset which did not show any significant issues.  There is no excess of power in the Fourier Transform from 1 to 12 hours has a 3 $\sigma$ detection limit of $\sim$15 ppm.  

\begin{figure}
\begin{center}
\includegraphics[scale=0.9]{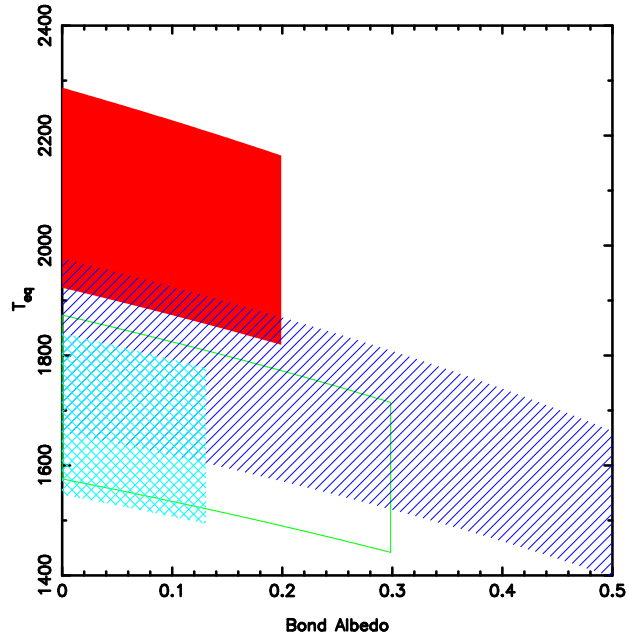}
\end{center}
\caption[Dayside Temperature of new Planets]{The dayside temperatures and 1 $\sigma$ upper bounds on the Bond albedo for {\it Kepler}-4b (hatched), {\it Kepler}-5b (solid), {\it Kepler}-6b (outside) and {\it Kepler}-7b (cross hatched) based on the 1-$\sigma$ upper limits on the the depth of the occultation.}
\label{fig:koitemps}
\end{figure}

Most of the observed light from the planet in the {\it Kepler} bandpass is due to thermal emission, but we can use the depth of the occultation to derive an upper limit on the albedo of the planet.  The geometric albedo,
\begin{equation}
A_g=\frac{F_p}{F_{\star}}\frac{a^2}{R^2_p},
\end{equation}
is defined as the ratio of the planet's luminosity at full phase to the luminosity from a Lambertian disk, where $R_p$ is the planetary radius. This allows us to place  boundaries on the dayside temperature of the planet as estimated by the equilibrium temperature given by Equation~\ref{eq:teq}.  Figure~\ref{fig:koitemps} shows the boundaries for {\it Kepler}-4b, 5b, 6b and 7b where we have calculated $T_{\rm eq}$ for $f = ${1,2} and $A_{\rm B}$ from zero to the 1-$\sigma$ upper limit.  The Bond albedo is estimated from the geometric albedo by assuming a Lambertian scatter as described in \citet{row06}.  Due to the relatively short periods of the first four planets discovered by {\it Kepler}, it not surprising that all of them have temperatures greater than 1500 K.  The upper limits show that Kepler-5b, 6b and 7b are less reflective than Jupiter and likely quiet dark, consistent with predictions and observations of other hot extrasolar planets.  Additional photometry of these planets will provide special opportunities to study planets in extreme conditions.

\acknowledgments
Funding for this Discovery mission is provided by NASA's Science Mission Directorate.  We would like to thank D. Fischer, A. Prsa and everyone that has contributed to the Kepler mission.

\begin{deluxetable}{lcc}
\tabletypesize{\scriptsize}
\tablewidth{0pc}
\tablenum{2}
\tablecaption{System Parameters for \koicur \label{tab:parameters}}
\tablehead{\colhead{Parameter}	& 
\colhead{Value} 		& 
\colhead{Notes}}
\startdata
Orbital period $P$ (d) & \koicurLCP & A \\ Midtransit time $E$ (HJD) &
\koicurLCT & A \\ Scaled semimajor axis $a/\rstar$ & \koicurLCar & A
\\ Scaled planet radius \rpl/\rstar & \koicurLCrprstar & A \\ Impact
parameter $b \equiv a \cos{i}/\rstar$ & \koicurLCimp & A \\ Orbital
inclination $i$ (deg) & \koicurLCi & A \\ Orbital semi-amplitude $K$
(\ms) & \koicurRVK & A,B \\ Orbital eccentricity $e$ & $< 0.024$ &
A,B,G \\ Center-of-mass velocity $\gamma$ (\ms) & \koicurRVgamma & A,B
\\
\sidehead{\em Observed stellar parameters}
Effective temperature \teff\ (K)                & \koicurSMEteff        & C     \\
Spectroscopic gravity \logg\ (cgs)              & \koicurSMElogg        & C     \\
Metallicity \feh                                & \koicurSMEfeh         & C     \\
Projected rotation \vsini\ (\kms)               & \koicurSMEvsin        & C     \\
Mean radial velocity (\kms)                     & \koicurRVmean         & B     \\
\sidehead{\em Derived stellar parameters}
Mass \mstar (\msun)                             & \koicurYYmlong        & C,D   \\
Radius \rstar (\rsun)                           & \koicurYYrlong        & C,D   \\
Surface gravity \loggstar\ (cgs)                & \koicurYYlogg         & C,D   \\
Luminosity \lstar\ (\lsun)                      & \koicurYYlum          & C,D   \\
Age (Gyr)                                       & \koicurYYage          & C,D   \\
\sidehead{\em Planetary parameters}
Mass \mpl\ (\mjup)                              & \koicurPPm            & A,B,C,D       \\
Radius \rpl\ (\rjup, equatorial)                & \koicurPPr            & A,B,C,D       \\
Density \rhopl\ (\gcmc)                         & \koicurPPrho          & A,B,C,D       \\
Surface gravity \loggpl\ (cgs)                  & \koicurPPlogg         & A,B,C,D       \\
Orbital semimajor axis $a$ (AU)                 & \koicurPParel         & E     \\
Equilibrium temperature \teq\ (K)               & \koicurPPteq          & F
\enddata
\tablecomments{\\
A: Based on the photometry.\\
B: Based on the radial velocities.\\
C: Based on a MOOG analysis of the FIES spectra.\\
D: Based on the Yale-Yonsei stellar evolution tracks.\\
E: Based on Newton's version of Kepler's Third Law and total mass.\\
F: Assumes Bond albedo = 0.1 and complete redistribution.\\
G: 1 sigma upper limit
}
\end{deluxetable}


\begin{thebibliography}{otherstuff}
\bibitem[Bakos et al.(2009)]{bak09} Bakos, G. \'A., et al. 2009, \apj, in press (arXiv:0901.0282v1)
\bibitem[Batalha et al.(2010)]{bat10} Batalha, et al. 2010, \apj, this issue
\bibitem[Borucki et al.(2010a)]{bor10a} Borucki, W. J., et al. 2010, Science, TBD
\bibitem[Borucki et al.(2010b)]{bor10b} Borucki, W. J., et al. 2010, \apj, this issue
\bibitem[Caldwell et al.(2010)]{cal10} Caldwell, D. A., et al. 2010, \apj, this issue
\bibitem[Claret(2000)]{cla00} Claret, A. 2000 \aap, 363, 1081
\bibitem[Daemgen et al.(2009)]{dae09} Daemgen, S., et al. 2009, \aap, 498, 567
\bibitem[Dunham et al.(2010)]{dun10} Dunham, E. W., et al. 2010, \apj, this issue
\bibitem[Ford(2005)]{for05} Ford, E.B. 2005, \apj, 129, 1706
\bibitem[Gautier et al.(2010)]{gau10} Gautier, T. N., et al. 2010, \apj, this issue
\bibitem[Gilliland et al.(2010)]{gil10} Gilliland, R., et al. 2010, \pasp, submitted
\bibitem[Haas et al.(2010)]{has10} Haas, M. et al. 2010, \apj, this issue
\bibitem[Jenkins et al.(2010)]{jen10} Jenkins, J. J., et al. 2010, \apj, this issue
\bibitem[Koch et al.(2010)]{koc10} Koch, D. G., et al. 2010, \apj, this issue
\bibitem[Latham et al.(2010)]{lat10} Latham, D. W., et al. 2010, \apj, this issue
\bibitem[Mandel \& Agol(2002)]{man02} Mandel, K., \& Agol, E. 2002, \apj, 580, L171
\bibitem[Morel(1997)]{mor97} Morel, P. 1997, \aap, 124, 597
\bibitem[P\'al et al.(2008)]{pal08} P\'al, A., et al. 2008, \apj, 680, 1450
\bibitem[P\'al et al.(2009)]{pal09} P\'al, A., et al. 2009, arXiv:0908.1705v2
\bibitem[Press et al.(1992)]{pre92} Press, W. H., Teukolsky, S. A., Vetterling, W. T., 
\& Flannery, B. P. 1992, Numerical Recipes in Fortran 77, Second Edition, Cambridge University Press, 678
\bibitem[Rowe et al.(2006)]{row06} Rowe, J. F., et al. 2006, \apj, 646, 1241
\bibitem[Walker et al.(2008)]{wal08} Walker, G. A. H., et al. 2008, \aap, 482, 691
\bibitem[Valenti \& Piskunov(1996)]{Valenti:96} Valenti, J.~A., \& Piskunov, N. 1996 \aaps, 118, 595
\bibitem[Yi et al.(2001)]{Yi:01} Yi, S.~K., Demarque, P., Kim, Y.-C., Lee, Y.-W., Ree, C.~H.,  Lejeune, T., \& Barnes, S. 2001, \apjs, 136, 417
\end{thebibliography}
\end{document}